\documentclass{appolb}

\newcommand{\sqrtsNN}{\mbox{$\sqrt{\mathrm{s}_{_{\mathrm{NN}}}}$}}
\newcommand{\axi}{$\overline{\Xi}^+$}
\newcommand{\xim}{$\Xi^-$}
\newcommand{\alam}{$\overline{\Lambda}$}
\newcommand{\lam}{$\Lambda$}
\newcommand{\ks}{$\mathrm{K}^{0}_{\mathrm S}$}

\newcommand{\ppt}{$p_{\rm T}$}

\def \pp    {$p + p$ }

\usepackage{epsfig}

\usepackage{amssymb} 

\usepackage{lineno} 

\begin{document}
\title{Measurements of \ks, \lam\ and $\Xi$ from Au+Au collisions at \sqrtsNN\ =~7.7, 11.5 and 39 GeV in STAR
\thanks{Presented at Strangeness in Quark Matter, Krak\'ow, Poland, September 18-24, 2011}
}
\author{Xianglei Zhu (for the STAR Collaboration)
\address{Department of Engineering Physics, Tsinghua University, Beijing 100084, China}
}
\maketitle


\begin{abstract}

We report on the measurements of \ks, $\Lambda$ and
$\Xi$ spectra at mid-rapidity ($|y|<0.5$) in the most central (0-5\%) Au+Au collisions at
\sqrtsNN\ =\ 7.7, 11.5 and 39 GeV from the STAR experiment. The extracted yields and the corresponding data from Pb+Pb collisions measured by the NA49 and CERES experiments at SPS are consistent. The \lam, \alam, \xim and \axi to $\pi$ ratios agree well with the predictions from a statistical hadronization model at all three energies.

\end{abstract}

\section{Introduction}

The enhanced production of strange hadrons in A+A with respect to p+p collisions has been 
suggested as a signature of Quark-Gluon Plasma (QGP) formation in these collisions \cite{raf82}. 
Until now, strangeness has been extensively measured in many experiments at different accelerator facilities \cite{e896,E802,e895,e917,e891,Antinori:2004ee,na49prc,Anticic:2009ie,ceres,Abelev:2007xp,Adams:2006ke,Abelev:2008zk,Adler:2002uv,Adams:2003fy,starprc83,phenix_lambda,lhc}. In these experiments, substantial strangeness enhancement has indeed been observed, especially for multi-strange hyperons. However, it is also observed that strangeness enhancement tends to increase toward lower energies. Generally, the yields of strange hadrons in nuclear collisions are close to those expected from statistical hadron gas models assuming a grand canonical ensemble \cite{becattini_prc,pbmnpa,pbm_shm}. According to canonical statistical models, strangeness enhancement in nuclear collisions can also be due to canonical suppression of strangeness production in \pp\ collisions \cite{Redlich:2001kb}. At SPS energies, there is a significant disagreement in strangeness measurement between NA49 and NA57 experiments \cite{Antinori:2004ee,na49prc,Anticic:2009ie}. The precise measurement of strangeness production in heavy ion collisions at SPS energy range will certainly lead to better understanding of strangeness enhancement/production mechanism in nuclear collisions. 

During the first phase of the Beam Energy Scan (BES) program in 2010, STAR has collected high statistics Au+Au data at \sqrtsNN\,\,=~7.7, 11.5 and 39~GeV, which allows high precision strangeness measurements at these energies. In STAR, strange hadrons are reconstructed with their secondary TPC tracks through the topology of their weak decay channels, \ks$\,\,\rightarrow \pi^+\pi^-$, $\Lambda\rightarrow p\pi$ and charged $\Xi\rightarrow\Lambda\pi$ \cite{Adler:2002uv,Adams:2003fy}. This analysis is based on about 5, 12 and 14 million minimum bias Au+Au events at \sqrtsNN\ = 7.7, 11.5 and 39 GeV respectively. 
In this paper, we will focus on the collision energy dependance and present the strangeness data from the most central (0-5\%) collisions at three BES energies. 

\section{Results}

\begin{figure} [h]
\centering
\includegraphics[width=0.9\textwidth]{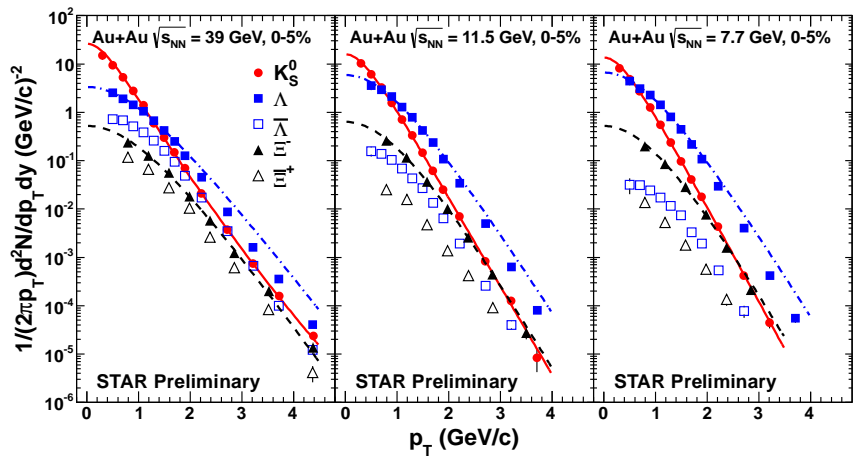}
\caption{Transverse momentum spectra of \ks, \lam(\alam), \xim(\axi)
at $|y|<0.5$ from the most central (0-5\%) Au+Au collisions at \sqrtsNN
= 39, 11.5 and 7.7 GeV. The solid, dot-dashed and dashed lines are the Levy (for \ks) and Maxwell-Boltzmann (for \lam\ and \xim) functions fitted to the corresponding spectra. Errors are statistical only. } \label{fig1}
\end{figure}

After correcting the raw spectra for geometrical acceptance and reconstruction efficiencies, we get the \ppt\ spectra of \ks, \lam(\alam), \xim(\axi) at mid-rapidity ($|y|<0.5$) for the most central Au+Au collisions (0-5\%) at \sqrtsNN\ = 7.7, 11.5 and 39 GeV, as shown in Fig. \ref{fig1}. The \lam(\alam) spectra have been corrected for the feed-down contributions from $\Xi$ and $\Xi^0$ weak decays. We assume $\Xi^0$ has the same spectra as the $\Xi$'s. The feed-down from $\Omega$ is neglected in this analysis. The percentage of the feed-down contribution in the raw \lam(\alam) spectra depends on the \lam(\alam) topological selection cuts, and decreases with the increase of \ppt. With the current cuts, the overall feed-down contributions are 22\%~(31\%), 18\%~(34\%) and 14\%~(47\%) for \lam~(\alam) at \sqrtsNN\,=\,\,39, 11.5 and 7.7 GeV, respectively.  

\begin{figure} [h]
\begin{minipage}[t]{0.35\textwidth}
\vspace{36pt}
\centering
\includegraphics[width=\textwidth]{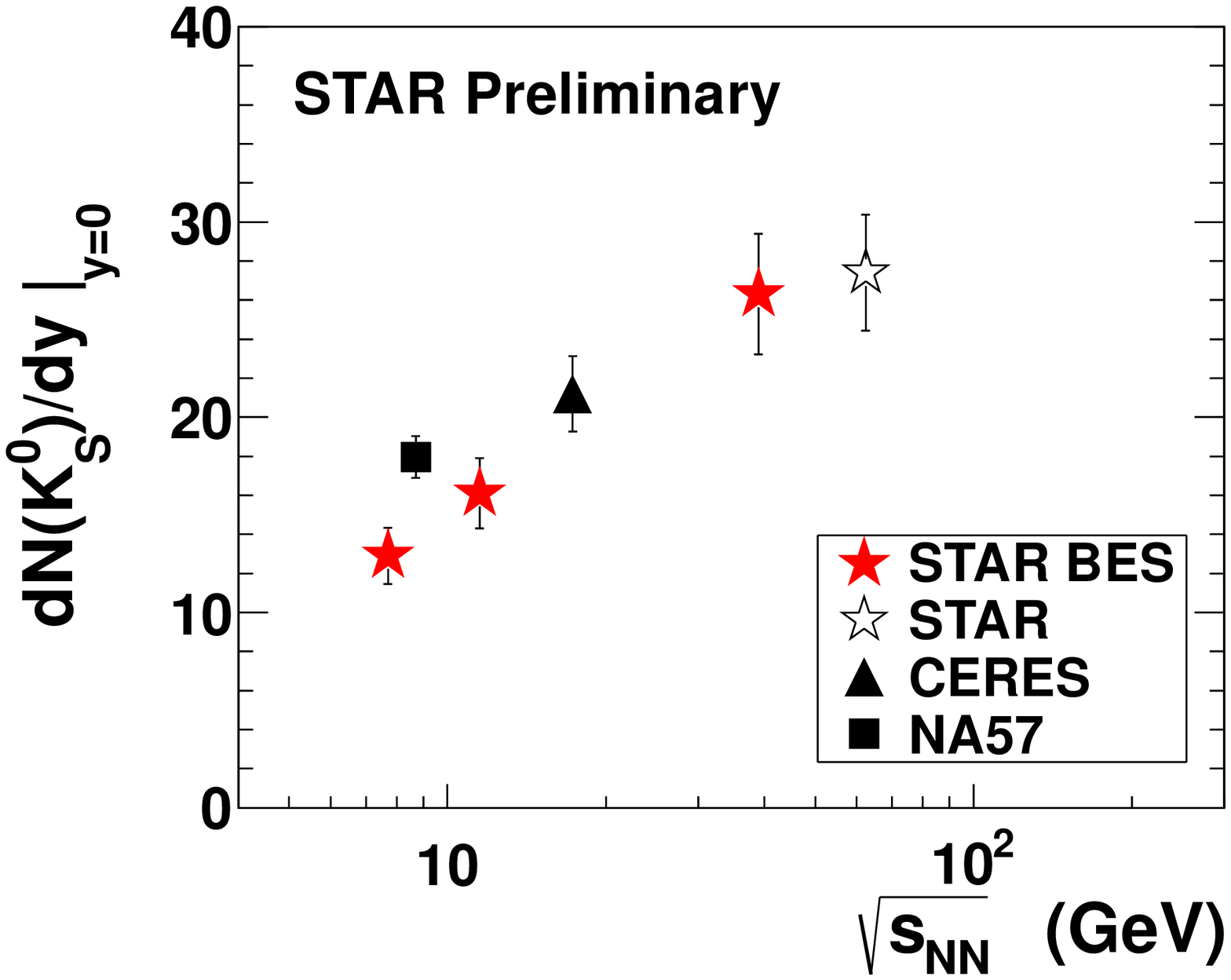}
\end{minipage}
\begin{minipage}[t]{0.65\textwidth}
\vspace{0pt}
\centering
\includegraphics[width=\textwidth]{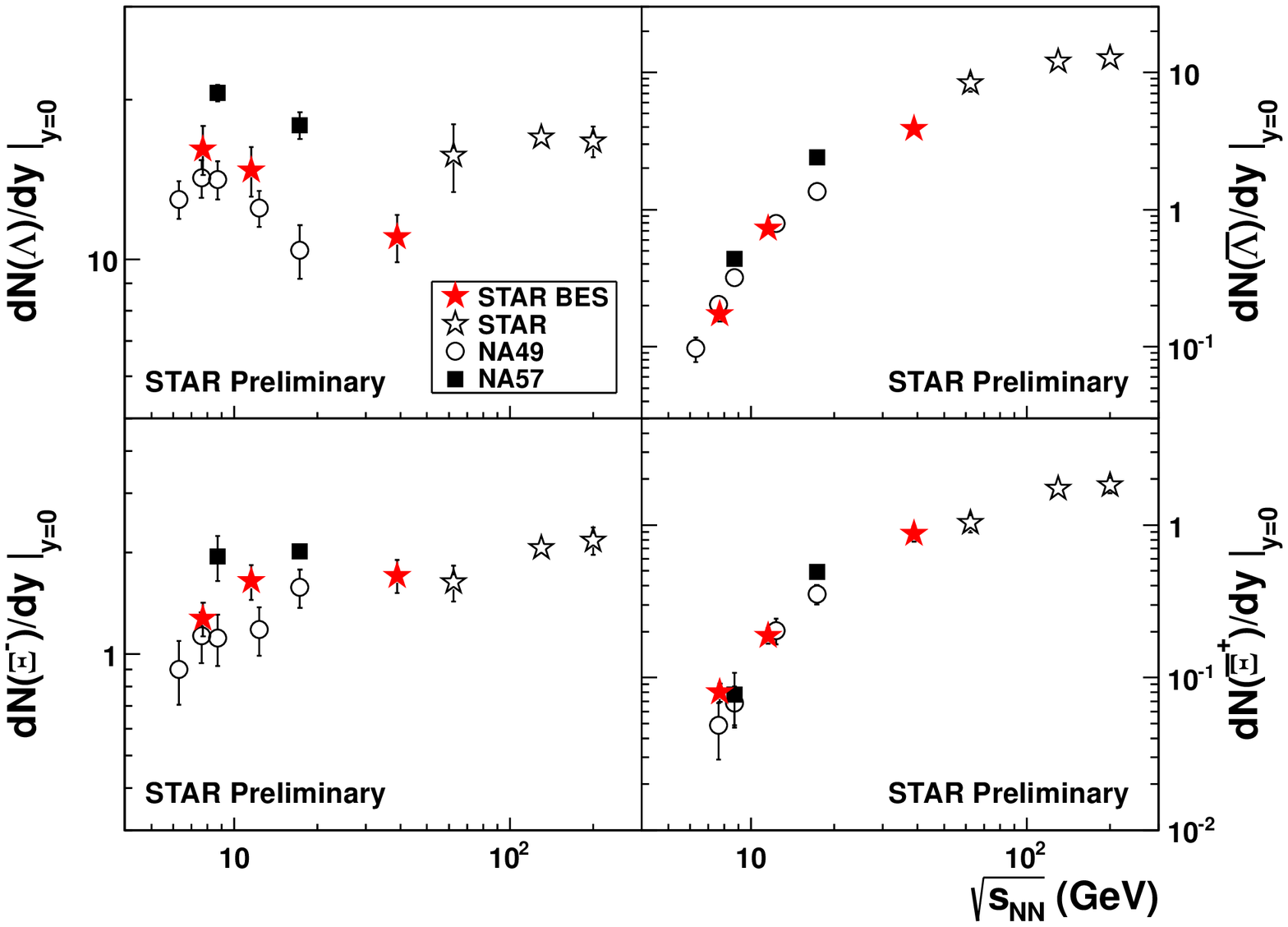}
\end{minipage}
\caption{The strange hadrons $dN/dy$ at mid-rapidity in the most central Au+Au collisions as a function of \sqrtsNN. Also shown are the results from STAR experiments at higher energies \cite{Adams:2006ke,Adler:2002uv,Adams:2003fy,starprc83}, Pb+Pb collisions from NA57 \cite{Antinori:2004ee} and NA49 \cite{na49prc}, and Pb+Au collisions from CERES \cite{ceres}. The rapidity ranges are $|y|<0.5$ for STAR and NA57, $|y|<0.4$ for NA49 \lam(\alam) and $|y|<0.5$ for NA49 \xim(\axi). CERES data are the extrapolated values based on the measurements at backward rapidity.} \label{fig2}
\end{figure}

In order to extract the total yield, the \ppt\ spectra have to be extrapolated to the unmeasured \ppt\ range with a function that is fit to the measured data. The \ks\ spectra can be well described with a Levy function \cite{Wilk:1999dr}, as shown in Fig. \ref{fig1}. The measured $\Lambda$ and $\Xi$ spectra can be extrapolated at low \ppt\ with the Maxwell-Boltzmann function ($1/p_{\rm T}d^2N/dp_{\rm T}dy \propto m_{\rm T}e^{-m_{\rm T}/T}$) or $m_{\rm T}$-exponential function ($1/p_{\rm T}d^2N/dp_{\rm T}dy \propto e^{-m_{\rm T}/T}$). Figure \ref{fig1} shows the corresponding Maxwell-Boltzmann function fitting curves for \lam\ and \xim (the fitting range is  \ppt$\,<1.6$ GeV/$c$ for \lam\ and \ppt$\,<2.6$ GeV/$c$ for \xim). The fitting can also be done with $m_T$-exponential function. The difference between the extrapolations with two different fitting functions is considered as the systematic error due to the extrapolation to unmeasured \ppt\ range, their values are about 3\% for \ks\,, 1\% for \lam(\alam) and 3\% for \xim(\axi). In fact, the main source of systematic uncertainties in the yield resides in the reconstruction efficiencies due to imperfect detector simulations. A previous estimation shows that this uncertainty is less than 10\% \cite{starprc83}, we take a value of 10\% in this analysis.

Figure \ref{fig2} shows the collision energy dependence of the particle yield ($dN/dy$) at mid-rapidity for
\ks, \lam, \alam, \xim\ and \axi\ from the most central (0-5\%) Au+Au collisions, compared to corresponding data from NA49, NA57 and CERES, as well as the STAR data at higher energies. The NA57 and NA49 data are from the most central Pb+Pb collisions, their data have been re-scaled according to the estimated numbers of wounded nucleons, $\left<N_W\right>$. The scale factor is $N_{part}/\left<N_W\right>$, where $N_{part}$ is the number of participants in the most central (0-5\%) Au+Au collisions in STAR. A Monte Carlo Glauber model estimation gives $N_{part}$\,= $337.4\pm2.3$, $338.4\pm2.1$ and $341.8\pm2.3$ at \sqrtsNN\,=\,7.7, 11.5 and 39 GeV. For simplicity, a $N_{part}$ value of 338 is used in the scale factor. Figure \ref{fig2} shows that STAR BES data lie in a trend with the corresponding data from higher energies, though there seems to be a non-monotonic \sqrtsNN\ dependence in the \lam\ $dN/dy$. While STAR BES data and NA49 data are consistent in general, the NA57 data are significantly higher except the \axi\ yield at \sqrtsNN\,=\,8.7\,\,GeV. The CERES \ks\ data seems in line with STAR BES data as well. It should be noted that the $\phi$ measurements in STAR BES also show consistent results with NA49 \cite{xiaoping}.
\begin{figure} [h]
\centerline{\includegraphics[width=0.5\textwidth]{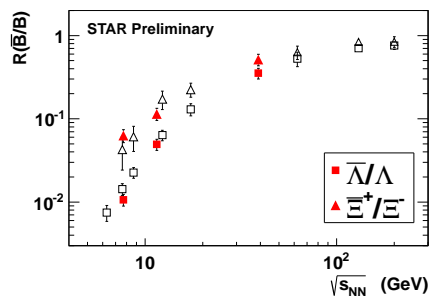}}
\caption{The \sqrtsNN\,  dependence of $\overline{\Lambda}/\Lambda$ and $\overline{\Xi}^+/\Xi^-$ ratios at mid-rapidity in the most central Au+Au collisions from STAR and Pb+Pb collisions from NA49. Solid symbols are STAR BES data; open symbols are STAR data at higher energies ($>\,$62\,GeV), and NA49 data at SPS energy range.} \label{fig3}
\end{figure}

Figure \ref{fig3} shows the anti-baryon to baryon ratios ($\overline{B}/B$) from STAR BES and comparisons to STAR higher energies and NA49 data. It seems that STAR BES data and NA49 data are consistent and in the published energy dependence trend. The $\overline{\Xi}^+/\Xi^-$ ratio are much larger than $\overline{\Lambda}/\Lambda$ at lower energies, which is consistent with the predictions from the thermal-statistical models \cite{becattini_prc,pbm_shm,Redlich:2001kb}. In these models, $\ln(\overline{B}/B)$ is proportional to the difference between strangeness numbers of $\overline{B}$ and $B$. And the coeffcient of proportionality ($\mu_S/T$) becomes more important at lower energies due to the large $\mu_B$ and the zero initial strangeness number.

\begin{figure} [h]
\centerline{\includegraphics[width=0.75\textwidth]{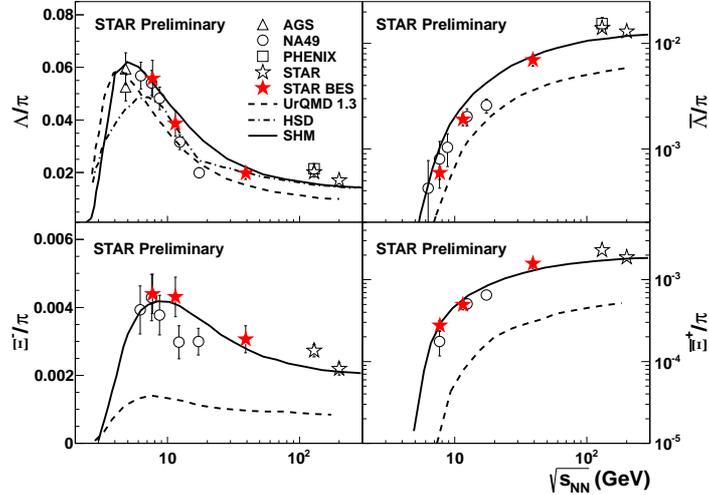}}
\caption{The $dN/dy$ of \lam , \alam, \xim\ and \axi at mid-rapidity divided by that of pions ($1.5(\pi^++\pi^-)$) in central Au+Au and Pb+Pb collisions as a function of \sqrtsNN. The $dN/dy$ of pions in STAR BES is taken from \cite{bes_pid}. Also shown are AGS \cite{e896,E802,e917,e891}, NA49 \cite{na49prc}, PHENIX \cite{phenix_lambda,phenix_pid} and STAR \cite{Adams:2006ke,Adler:2002uv,Adams:2003fy,starpid_130,starpid_200} data, as well as predictions from two hadronic transport models: HSD and UrQMD (v1.3) \cite{hsdurqmd} and a statistical hadron gas model (SHM) \cite{pbm_shm}. } \label{fig4}
\end{figure}

Figure \ref{fig4} shows the \sqrtsNN\ dependence of the ratios of \lam, \alam, \xim\ and \axi\ mid-rapidity $dN/dy$ to that of pions in STAR BES, as well as the existing data from various experiments at different energies and the predictions from hadronic transport models (UrQMD v1.3 and HSD). Though the hadronic models seem to reproduce the $\Lambda/\pi$ data, the default UrQMD (v1.3) fails in reproducing the $\Xi^-/\pi$ and \axi$/\pi$ ratios due to less $\Xi$ yield in the model. The data is also compared with the calculations from a statistical hadron gas model (SHM). This model is based on a grand canonical ensemble and assumes chemical equilibrium. And the parameters $T_{ch}$ and $\mu_{B}$ in the model are obtained from a smooth parametrization of the original fitting parameters to the mid-rapidity particle ratios from heavy ion experiments at SPS and RHIC. Figure \ref{fig4} shows that SHM model predictons agree well with data across the whole energy range from AGS to top RHIC energies. 

\section{Summary}

In summary, we presented the transverse momentum spectra and $dN/dy$ of \ks, \lam\ and $\Xi$ at mid-rapidity, as well as the $\overline{B}/B$ and strange hadron to pions ratios, from central Au+Au collisons at \sqrtsNN = 7.7, 11.5 and 39 GeV at RHIC. The strange hadrons $dN/dy$ from STAR BES and the corresponding NA49 and CERES data seem consistent in general, and are much lower than the data from NA57. The $\overline{B}/B$ ratios are in the published energy dependence trend. The ratios of \lam, \alam, \xim\ and \axi\ to pions at mid-rapidity agree well with the predictions of statistical hadronization model.
\\
\\
{\bf Acknowledgments:} X. Zhu thanks the support by National Natural Science Foundation of China (Grant Nos. 10905029, 11035009) and the Foundation for the Authors of National Excellent Doctoral Dissertation of P.R. China (FANEDD) (No. 201021).

\end{document}